\documentclass[aip,pof,reprint,superscriptaddress]{revtex4-1}
\pdfoutput=1
\usepackage[pdftex,bookmarks,colorlinks]{hyperref}
\usepackage[pdftex]{graphicx}
\usepackage{amssymb,amsfonts,amsmath,graphics}
\DeclareGraphicsExtensions{.pdf,.png,.jpg}
\usepackage{pbox}
\usepackage[normalem]{ulem}
\usepackage{color}
\usepackage{enumerate}
\begin{document}
\title{Capillary droplets on Leidenfrost micro-ratchets}

\author{\'Alvaro G. Mar\'in}\altaffiliation{Now at Institute of Fluid Mechanics and Aerodynamics, Bundeswehr Univ., Munich, Germany.}
\email{a.marin@unibw.de}
\affiliation{Physics of Fluids, Mesa+, Faculty of Science and Technology, University of Twente, The Netherlands.}

\author{Daniel Arnaldo del Cerro}
\affiliation{Applied Laser Technology, Faculty of Engineering Technology, University of Twente, The Netherlands.}

\author{Gertwillem R.B.E. R\"omer}
\affiliation{Applied Laser Technology, Faculty of Engineering Technology, University of Twente, The Netherlands.}

\author{B. Pathiraj}
\affiliation{Applied Laser Technology, Faculty of Engineering Technology, University of Twente, The Netherlands.}

\author{Albertus Huis in 't Veld}
\affiliation{Applied Laser Technology, Faculty of Engineering Technology, University of Twente, The Netherlands.}
\author{Detlef Lohse}
\email{detlef.lohse@utwente.nl}
\affiliation{Physics of Fluids, Mesa+, Faculty of Science and Technology, University of Twente, The Netherlands.}

\begin{abstract}

Leidenfrost ratchets are structures with the ability of transporting liquid droplets when heated over the critical Leidenfrost temperature. Once this temperature is reached, the droplet levitates over the surface and moves in the direction marked by the slope of the ratchet at terminal velocities around $10~cm/s$. Here we provide new experiments with micron-sized ratchets, which have been produced with picosecond pulse laser ablation. 
In the following work, we use a simple method to measure the thrust driving droplets of capillary size over the micro-ratchets. The mechanism responsible for the force acting on the drop on superheated ratchets has been recently under debate. We extend the recently proposed  ``viscous mechanism'' proposed by Dupeaux et al. [Europhys. Lett., 96, 58001 (2011)] to capillary droplets and find good agreement with our measurements.

\end{abstract}

\maketitle

\section{Introduction}

Ratchets are quite ubiquitous systems in many disciplines: In zoology, for example, the so-called ``Capillary-Ratchet''\cite{prakash2008surface} stands for an effect employed by animals to drink water. In sport disciplines as ski-touring, ratchet-like furry seal skin was traditionally used to climb uphill on snow with the skis. Nowadays this popular sport uses synthetic ``skin'' with improved properties. Recent research has also been devoted to non-equilibrium systems in which the symmetry of the system is broken through ratchets to force the dynamics in a particular direction, that is the case of new methods proposed to separate particles passively \cite{reguera2012entropic}. A classical example is that of the Feynman-Smoluchowski Ratchet, which caused strong debate and further research since its entropy production seems to surprisingly follow detailed fluctuation theorems\cite{granularratchet1,granularratchet2,Joubaud.Ratchet}.
Recently, Linke et al. \cite{Linke:Leidenfrost} found that a surface carved as a ratchet and heated over the Leidenfrost point is able to propel an evaporating droplet in a preferential direction. The Leidenfrost point is the temperature at which the vapor emanated from the droplet is able to sustain the droplet's own weight. As a consequence, a droplet deposited over a surface that has been heated above this temperature will levitate without friction \cite{Biance:2003}. The discovery of Linke et al. permits a better control on Leidenfrost droplets and allows for new systems in which drops are quickly and precisely transported to their destination hovering over ratcheted-heated surfaces.
\begin{figure}[h*]
\includegraphics[width=0.5\textwidth]{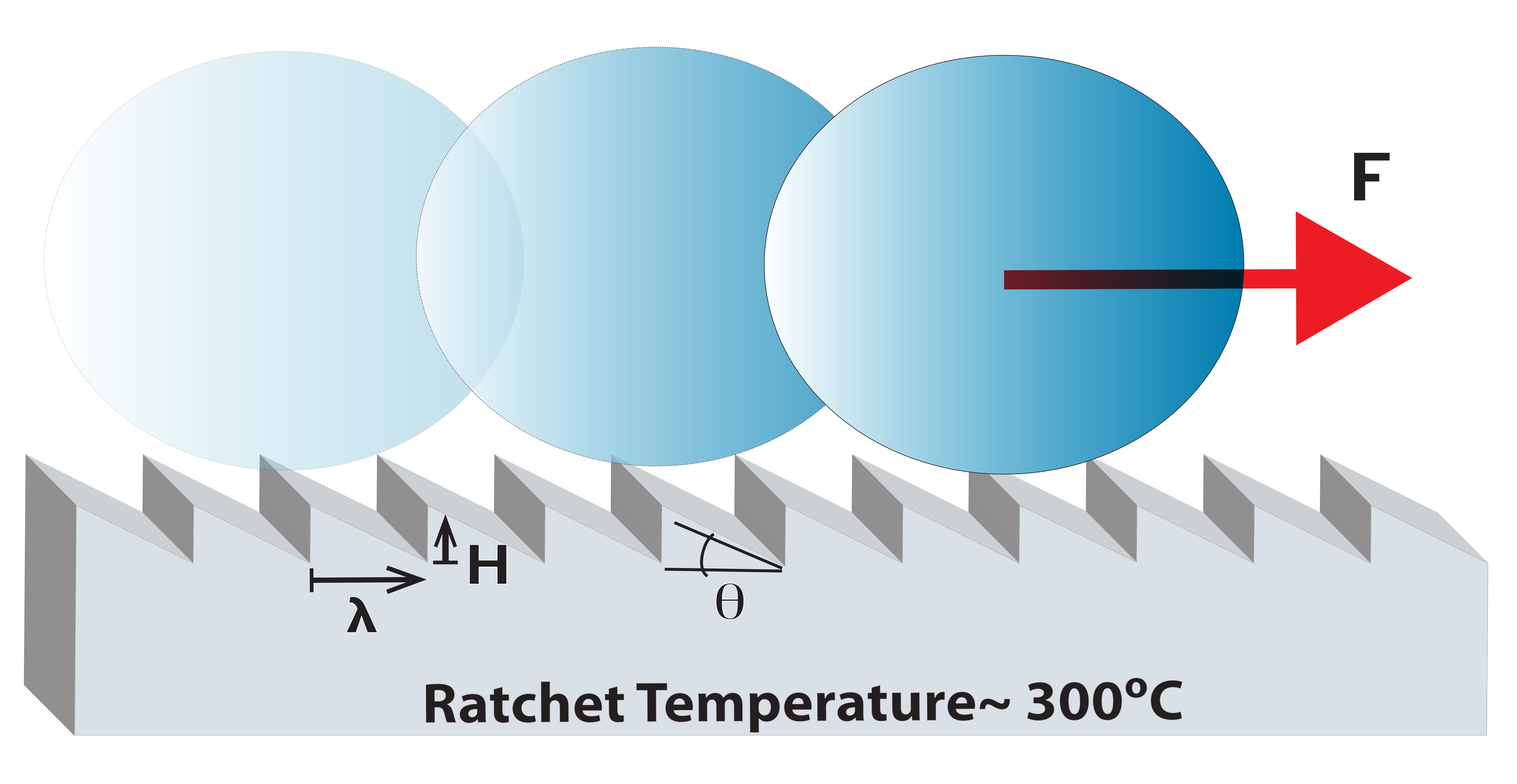}
\caption{Scheme of the Leidenfrost ratchet: an evaporating liquid drop is deposited on a heated metallic surface patterned with ratchet-like crenelations. The droplet will feel a force and will move in the indicated direction.}
\label{figure1}
\end{figure}
D. Qu\'er\'e, C. Clanet and coworkers, have intensively investigated the physics and hydrodynamics leading to the propulsion and friction found in the droplets \cite{lagubeau2011leidenfrost,Dupeux:2011,dupeux2011trapping}. The main idea of the mechanism behind the so-called \emph{Leidenfrost Ratchet} is that the vapor flow inside each triangular-shaped cavity is forced in a preferential direction. This flow then drives the vapor coming from the droplet towards the opening and therefore exerts a drag upon the droplet. The mechanism described here is the so-called ``viscous mechanism''. However also other mechanisms have recently been proposed in the literature and will be discussed along the paper. 

\begin{figure}[h*]
\includegraphics[width=0.5\textwidth]{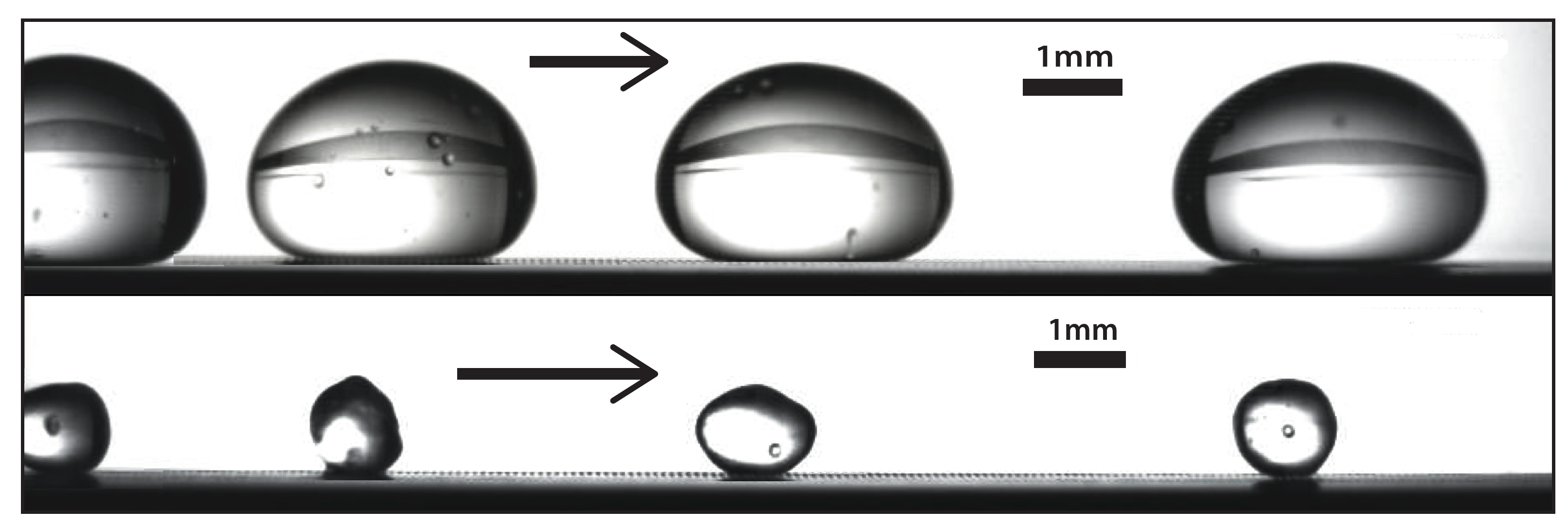}
\caption{Top: Large water droplet of 3 mm diameter on a micro-ratchet at different instants after deposition: 500, 1000, 1500 and 2000 ms. Note that the roughness of the ratchets ($\lambda \sim 140~\mu m$) is not visible at this scale. Bottom: Small capillary water droplet of 1 mm diameter on a microratchet at different instants after deposition: 50, 100, 150, and 200 ms. Note that the capillary droplet (bottom) travels the same distance 10 times faster than the larger one (top), (enhanced online).}
\label{drops}
\end{figure}

Different applications based on droplet transport have been proposed for the Leidenfrost Ratchets, however the lack of friction in these systems brings also a lack of control. To solve this situation, symmetrical crenelations can be used to increase the friction with the substrate and finally completely stop the droplet \cite{dupeux2011trapping}. Another way is the use of different ratchet geometries, for example circular ones, which can serve as Leidenfrost traps \cite{cousins2012ratchet}. Recently, another approach has been explored by using magnetic fields on Leidenfrost droplets of paramagnetic fluids as liquid oxygen \cite{LeidenfrostMagneto}.

\begin{figure}[h*]
\includegraphics[width=\textwidth]{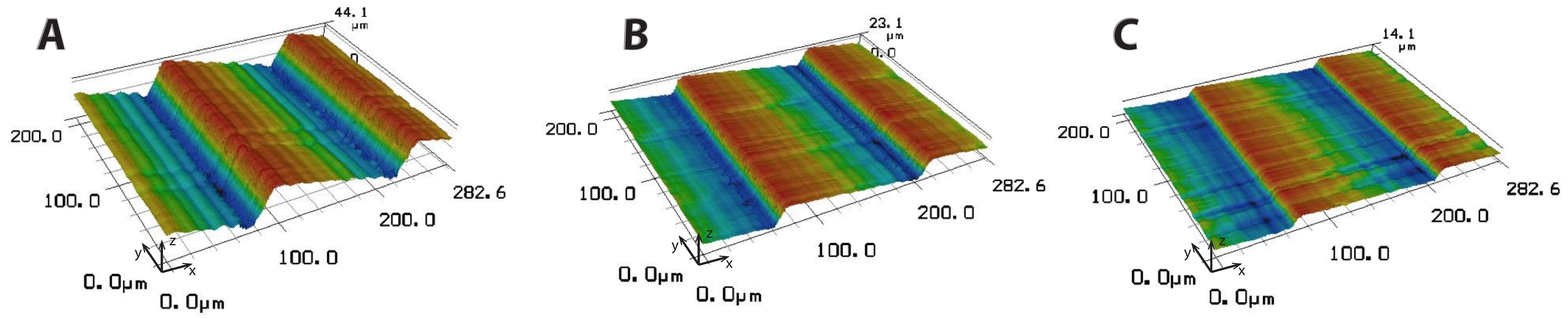}
\caption{Different ratchets employed in the study. They all keep the same ratchet length, but varying its depth, they therefore differ basically in the aspect ratio: a) $H/\lambda=0.23$ b) $0.11$ c) $0.07$. All images have been taken by confocal microscopy.}
\label{ratchets}
\end{figure}

The purpose of the current work is to test the different models for the propulsion of water capillary droplets on Leidenfrost ratchets by using micro-ratchets carved in stainless steel. The reduced size of the ratchets permits the use of smaller droplets sizes; the smallest droplet size reached is almost one order of magnitude larger than the typical ratchet size.

\section{Experimental set-up}
The micro-ratchets were produced by direct material removal using a picosecond pulsed laser source. Surface micro-patterning with picosecond laser pulses allows creating a well controlled topography on a variety of substrates, with a resolution typically in the micron range \cite{arnaldo2010,Meijer2002}. The laser source operates at a central wavelength of 1030 nm, and delivers Gaussian pulses below 10 ps with a beam quality parameter M$^2<$ 1.3. A second harmonic generation unit was employed for converting the central wavelength to 515 nm. A 103 mm  f-$\theta$ telecentric-lens was employed for the final beam focusing. The beam focus diameter was measured to be 16 $\mu$m. The resulting effective diameter, and the sufficient depth of focus which the lens provides, allow texturing large enough areas. Manipulation of the laser beam over the substrates was performed by a two mirror galvo scanner system, with a positioning accuracy of 1 $\mu$m. The ratcheted surfaces were created after a layer by layer process, removing thin slices of material at each step. A layer consisted of parallel laser tracks. The separation between the tracks was adjusted to reduce the resulting roughness. For the same purpose, the relative position of consecutive layers was shifted 50$\%$. The average power per pulse, and the spatial separation between pulses in a track, were adjusted to provide a fast laser ablation rate with limited thermal damage to the sample. 
In figure \ref{ratchets} we can see three different ratchet geometries engraved with the ultra-fast laser technique with three different cavity aspect ratios, defined as $H/\lambda$, where $H$ is the maximum vertical distance from top to bottom (\emph{y} axis in figure \ref{ratchets}) and $\lambda$ is the distance between the ratchet crests (\emph{x} axis in figure \ref{ratchets} and \ref{figure1}). The three ratchets have approximately a constant $\lambda\sim140~\mu m$, but different heights H, yielding different three aspect ratios: 0.23, 0.11 and 0.07 (figure \ref{ratchets}).

Droplets of purified water with diameters from 0.7 to 3 mm have been tested in the experiments. The droplets are deposited as gently as possible from needles of diameters ranging from 0.1 up to 1 mm. The desired liquid volume is injected through a syringe pump, then passing through the needle, sustained with a positioner which is located close enough to the heated surface so that the deposition is as smooth and reproducible as possible. The droplets are always injected from the same height.
The contact with the overheated surface however never comes without consequences. Big droplets upon contact show strong oscillations along their surface (same type as observed by Linke et al.\cite{Linke:Leidenfrost}) but smaller droplets bounce off the surface. The latter case of capillary droplets is specially interesting because the contact occurs in a matter of a few milliseconds. This seems to give enough time to the droplet to be propelled along the ratchet. The bouncing behavior consists then on a horizontal movement along the ratchet, combined with a vertical oscillatory bouncing which amplitude decreases as the droplet travels, dying in typical distances of a few millimeters. This behavior will have consequences in some of the measurements and will be discussed further below.

As soon as the droplet makes contact with the surface, it immediately moves along the ratchet following a fairly straight line. The drop's short voyage was filmed with a high-speed camera at typical recording speeds of 1000 fps, on 4 cm long ratchets. In order to calculate the physical parameters of each droplet trajectory, we will assume that the only drag experienced by the drop is that exerted by the surrounding air. Since the typical value of the Reynolds numbers is around 15, we will use Stoke's friction law to calculate the drag, which will depend linearly on the object's velocity. The validity of such assumptions will be confirmed below with an alternative method to measure the propelling force. The droplet trajectory will consequently satisfy the following ordinary differential equation:
\begin{equation}
m\frac{d^2x}{dt^2}=F_{R}-\beta\frac{dx}{dt},\label{Force}
\end{equation}
where $x$ is the droplet position along the ratchet, $m$ is the drop's mass, $F_R$ is the force that the ratchet exerts over the droplet (assumed to be independent of time, and on the droplet's position or velocity), and $\beta$ is the friction coefficient. Under this assumption, the droplet trajectory is then fit to the following function:
\begin{equation}
x(t)=a e^{-bt}+ct+d,
\label{xt}
\end{equation}
where $t$ is the time of flight, $x(t)$ is the position in time and $a$,$b$, $c$ and $d$ are interpolating parameters. Taking the derivative of this equation, we obtain the velocity:

\begin{equation}
v(t)=(v_o-v_T) e^{-\beta t/m}-v_T.
\label{vt}
\end{equation}
where $v_o$ is the initial velocity, $v_T$ the terminal velocity and $\beta$ the friction coefficient. 
Once the parameters $a$, $b$, $c$ and $d$ are found by fitting the experimental data, we can obtain the physical parameters in equation (\ref{vt}): $v_o$, $v_T$, $\beta$ and most importantly $F_R$. In the insert of figure \ref{veloplot} we can see how the fit works for a droplet trajectory. A common problem of the method is found for the fastest droplets, which need very long ratchets to reach their terminal velocity. However, as we can see in the insert of figure \ref{veloplot}, we can obtain a good estimation for the terminal velocity from the fit even with the fastest droplets.

This approach gives us then a simple but systematic way to calculate the force felt by the droplets in flight. However, we must stress that some assumptions as the linear drag have been made. With the intention of checking the validity of these assumptions, the force is measured following an alternative approach, which we call the \emph{inclined plane} method. This method consists on tilting the surface slightly until a droplet of a certain mass $m$ can no longer ``climb'' the plane (see supplementary material [reference here: "TiltedRatchet.mov"] for a video with a capillary droplet climbing successfully the ratchet but falling back once it reaches the end of the ratchet-patterned area). Since the maximum forces are of the order of a few $\mu N$, the highest tilting angles are in the order of a few degrees and therefore the errors are quite large due to the difficulty in determining the tilt of the surface precisely. 
Nonetheless, the \emph{inclined plane} measurements are intended only to validate the more precise \emph{velocity fit} method previously described.

\begin{figure}[h]
\includegraphics[width=0.5\textwidth]{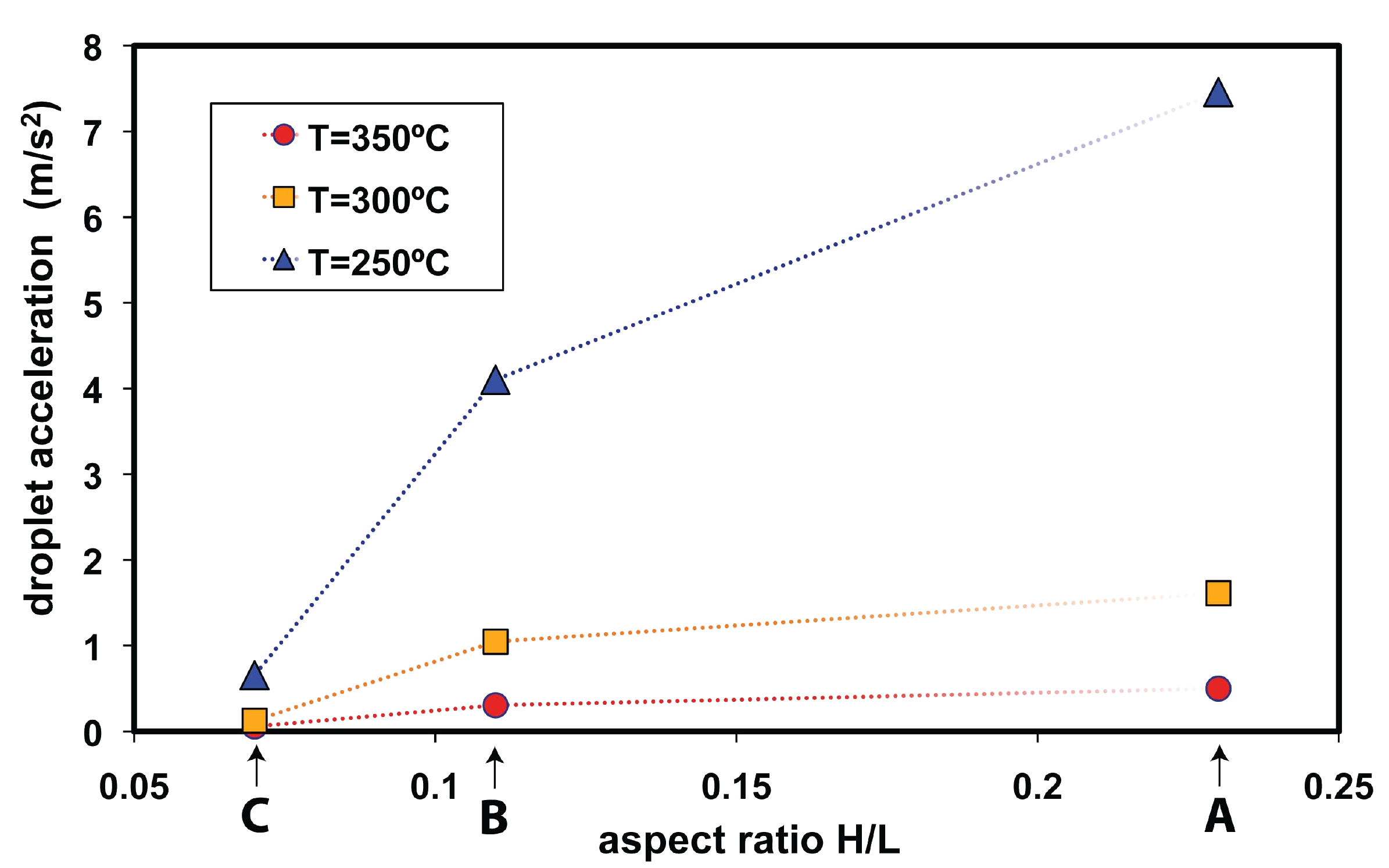}
\caption{Droplet acceleration for different ratchet temperatures and aspect ratios for drop volumes of $2~\mu l$ (ca. 1.6 mm diameter), measured as an average acceleration after a fixed traveling length of 4 cm. The dotted lines are only a guide to the eye. (enhanced online). }
\label{accVSaspect}
\end{figure}

\section{Experimental Results}

In order to find the optimal working conditions, different ratchet geometries and temperatures were tested. In figure \ref{accVSaspect} we can see the typical accelerations found for $2 \mu l$ droplets (ca. 1.6 mm in diameter) in microratchets with 3 different aspect ratios, from $0.07$, $0.11$, and $0.23$, and 3 different temperatures $250$, $300$, and $350^\circ$C. The dependence of the acceleration on the ratchet aspect ratio can be explained qualitatively by symmetry arguments, which will be developed more quantitatively later: the lower the aspect ratio, the more symmetric the structure will become and therefore the mechanism is less and less effective. Regarding the temperature, we observe that once the Leidenfrost temperature is achieved (in this case around $240^\circ$C) the acceleration decreases as the temperature is further increased. A similar trend has been observed by other authors \cite{Ok}. This fact can be easily understood if one assumes that the thrust is originated from the viscous shear that the moving vapor exerts under the droplet at its surface\cite{Dupeux:2011}. Consequently, as the temperature is increased beyond the Leidenfrost temperature, the average droplet-ratchet distance increases, the net force is reduced and therefore the acceleration is reduced. 

\begin{figure}[h]
\includegraphics[width=0.5\textwidth]{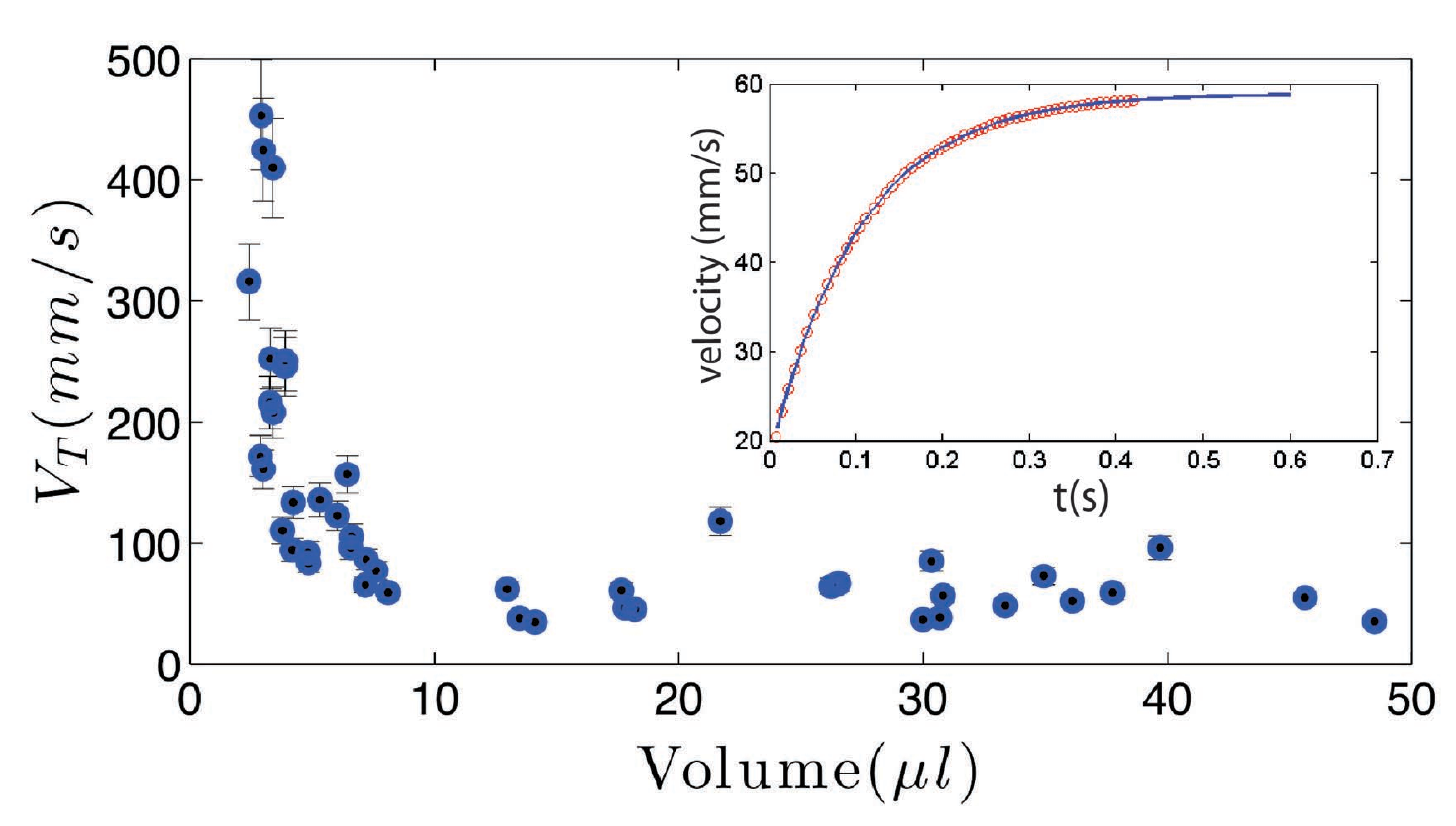}
\caption{Droplet velocity for different drop volumes. The velocities seem to stabilize around a 80mm/s for bigger droplets, but present larger and more fluctuating values for the smallest ones. Insert: Experimental droplet velocity as function of time. Circles represent experimental measurements. Continuous line represents the model fit used to determine the terminal velocity.}
\label{veloplot}
\end{figure}

In figure \ref{veloplot} we plot the droplets velocity as function of the volume. As the droplet size increases, their terminal velocity seem to stabilize to a constant value of $80~mm/s~(\pm 10~mm/s)$. This value is quite close to that found by Dupeux et al. \cite{Dupeux:2011} and Lagubeau et al. \cite{lagubeau2011leidenfrost} for droplets much larger than the typical ratchet length scale. In their work, when the droplet size becomes comparable to the ratchet length scale, the velocities went abruptly to zero. In our case, our smallest droplet is still about five times larger than the typical ratchet size and therefore the velocities do not go to zero but seem to considerably increase and show large fluctuations. The reason for such fluctuations lies in the natural oscillations experienced by the droplets when put in contact with the hot substrate. Smaller droplets tend to bounce off the substrate and bigger droplets display surface oscillations. In both cases this results in large perturbations at the bottom liquid-vapor interface, which is crucial for the development of the propelling force. This is the origin of the large dispersion of the data and cannot be accounted for an error estimation, which only includes measurement errors.

\begin{figure}[h]
\includegraphics[width=0.5\textwidth]{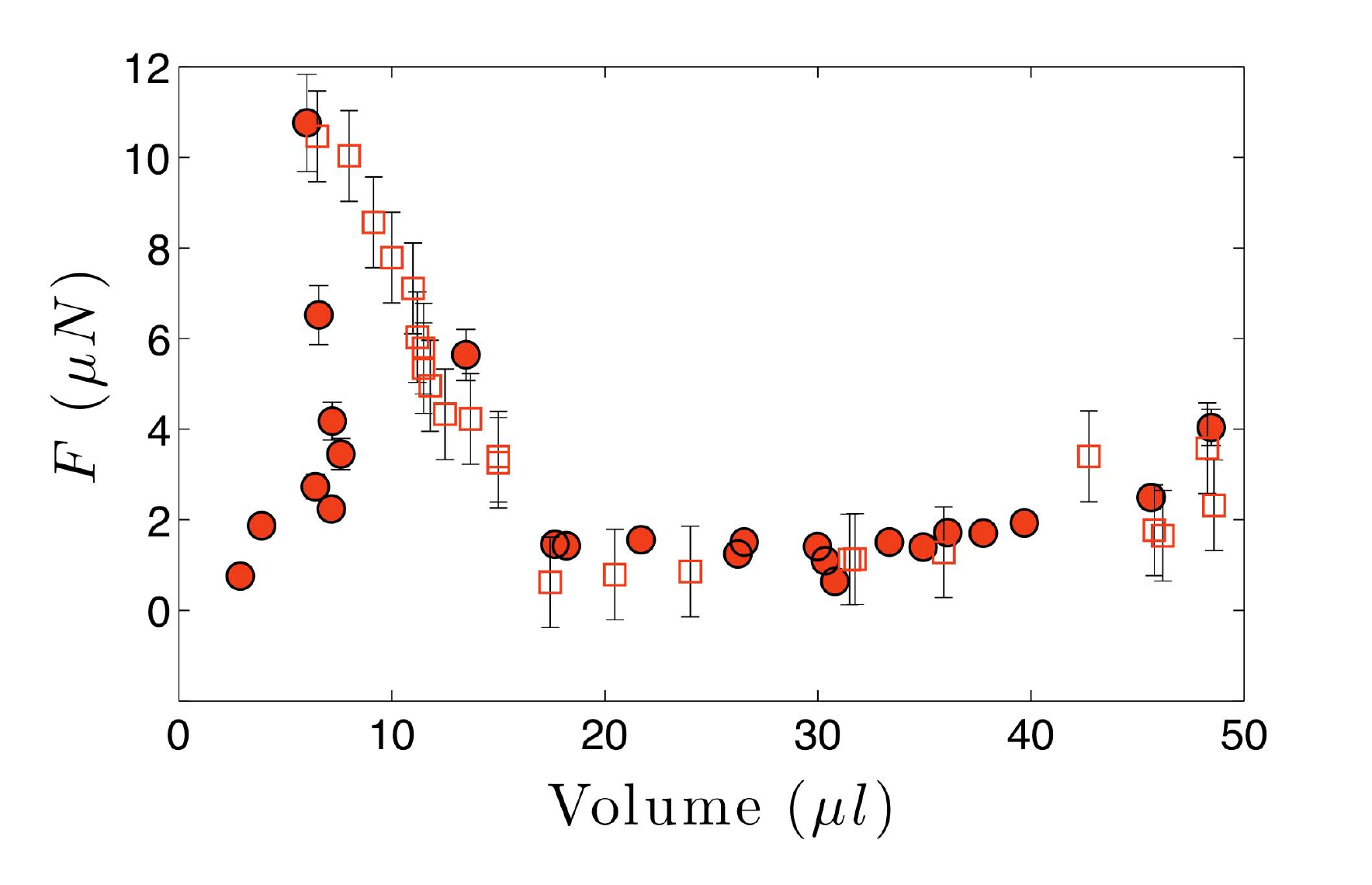}
\caption{Force measured for different drop volumes. Filled circles represent those measurements done with the \emph{velocity fit method} and those with open squares represent those done with \emph{inclined plane method}.}
\label{Fexp}
\end{figure}

In figure \ref{Fexp} we plot the force $F_R$ measured with both methods, \emph{velocity fit} and \emph{inclined plane}. Both methods seem to yield similar results, which confirms the assumptions made for the velocity fit method. For large droplets the force seems to decrease slowly with the drop size, as expected by the scaling laws proposed by Lagubeau et al. \cite{lagubeau2011leidenfrost} and Dupeux et al.\cite{Dupeux:2011}. However, as the droplet size reaches the capillary size, the force increases significantly in both types of measurements as the droplet size is reduced. The force then tends to decrease as the droplet size approaches the ratchet length. In figure \ref{Fexp} the data from the most pronounced bouncing droplets have been eliminated, which would yield even higher force values. Such high thrust observed on capillary droplets is indeed consistent with the terminal velocity data plotted in figure \ref{veloplot}. In order to give an explanation for the results we have to analyze the different models that have been proposed for the origin of the thrust of the Leidenfrost ratchet droplets.

\section{Scaling of the force}

Different mechanisms have been proposed for the thrust responsible for the drop movement. Lagubeau et al. \cite{lagubeau2011leidenfrost} proposed a simple mechanism based on vapor flow rectification and mass conservation, obtaining a scaling law $F \sim R^{3/2}$. Shortly later, new observations led to a more sophisticated mechanism: the vapor flow is indeed responsible of the thrust, but due to the viscous drag exerted over the drop's surface and not as a direct consequence of the momentum conservation\cite{Dupeux:2011}. This yields a force that scales with the volume of the droplet. Some other mechanisms have been proposed very recently, as that by W\"urger\cite{wurger2011}, in which a thermal creep generated from a huge thermal gradient along the ratchet surface can be responsible for the force felt by the droplet. Such mechanism yields a similar scaling with the drop size as that proposed by Lagubeau et al. \cite{lagubeau2011leidenfrost}, but temperature gradients of around $100~K/mm$ are needed to transport droplets at the speed observed in the experiments, which is a quite unrealistic situation when using thermally-conducting metals.
Recently, Hardt et al. \cite{hardt} performed numerical calculations solving the Boltzman equation for vapor molecules in the ratchet, confirming the suitability of the viscous mechanism and discarding other mechanisms as the thermal creep flow.
In the following we will further develop on the model by Dupeux et al. \cite{Dupeux:2011}, which gave the best results for our experiments. The idea behind the \emph{viscous mechanism} is the following: The vapor emanating from the bottom of the droplet flows preferentially into the deeper area of each ratchet, this gives the droplet a net shear in a preferential direction. 

\begin{figure}[h]
\includegraphics[width=0.8\textwidth]{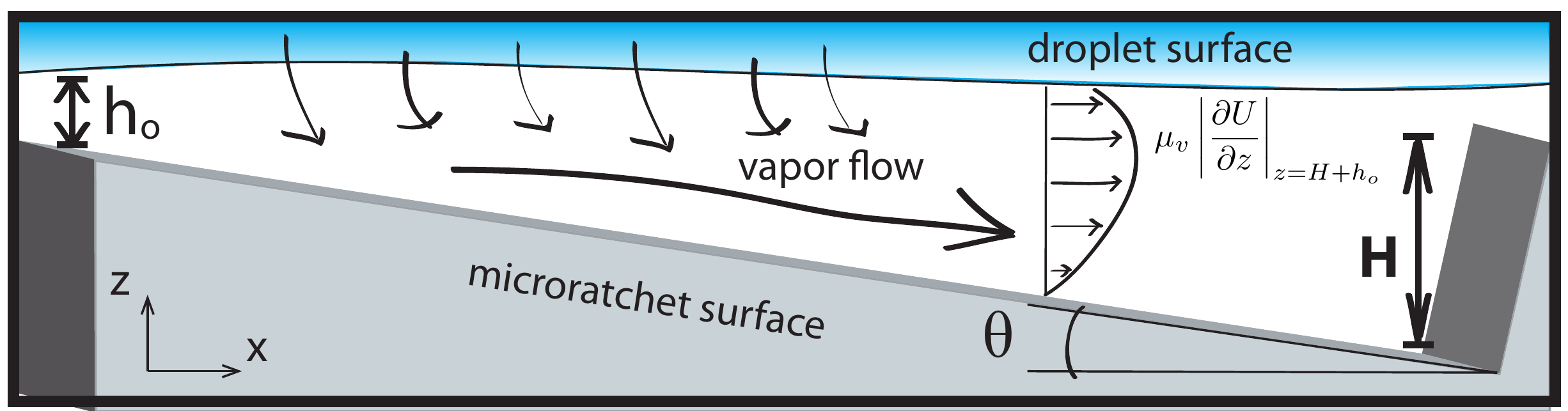}
\caption{Representation of the \emph{viscous mechanism} on a single microratchet: the vapor from the lower droplet surface penetrates into the ratchet, the ratchet geometry generates an horizontal pressure drop towards the right side, leading to a shear stress $\mu_v\partial U/\partial z$ at the droplets interface, which integrated over the contact area $\pi r_c^2 $ gives rise to a net thrust force towards the positive \emph{x}-axis.}
\label{sketch}
\end{figure}

Following the model presented by Biance et al. \cite{Biance:2003} we are able to estimate the vapour film thickness $h_o$ as a function of the drop's size, surface temperature, vapor viscosity and density, and thermodynamic properties. First, we make use of the energy conservation: a droplet on a hot plate will evaporate at a rate:
\begin{equation}
\dot m\sim\frac{k}{L}\frac{\Delta T}{h_o}\pi r_c^2,
\label{evaporationrate}
\end{equation}
where $k$ is the thermal conductivity, $L$ is the water latent heat of evaporation, $\Delta T/h_o$ is the temperature gradient from the top of the ratchet to the droplet surface, and $r_c$ is the contact radius. We will call contact area  that area which seems to be parallel to the surface, i.e. the area in which the heat transfer is effective. Several theoretical\cite{snoeijer2009maximum} and experimental studies \cite{burton2012geometry} have revealed the real shape of the lower interface of Leidenfrost droplets on hot flat plates, but since our model only aims to give a scaling law we will from now on take our lower interface as flat and discuss further on its implications and limitations. In the lubrication approximation for the vapor flow $\partial_xP=\mu_v\partial^2_zU$ and therefore $\Delta P/R\sim \mu_v U/h^2$. We can then write the vapor mass flow rate as: 
\begin{equation}
\dot m \sim \rho_v 2 \pi R h_o U \sim \rho_v \frac{h_o^3}{\mu_v}\Delta P.
\label{lubrication}
\end{equation}
where $\Delta P$ is the pressure imposed by the drop presence on the vapor layer and $\mu_v$ is the vapor viscosity. The weight of the droplet determines the contact radius and the balance of forces in the vapor layer.

We will first discuss the case of heavy drops, which shape is dominated by gravity: (a) the contact radius $r_c$ approximately equals the radius $R$ of the drop and (b) the pressure drop created by the droplet is proportional to its weight $\Delta P \sim \rho_l gH_{drop}$, where $H_{drop}$ is the droplet height, $g$ the gravitational acceleration and $\rho_l$ the liquid density. In this situation, combining equations (\ref{evaporationrate}) and (\ref{lubrication}) yields an expression for the film thickness for heavy drops $h_o^{(g)}$:

\begin{equation}
h_o^{(g)}\sim\left(\frac{k\Delta T}{L\rho_v}\frac{\mu_v}{\rho_l g H_{drop}}\right)^{1/4}R^{1/2}
\label{hheavy}
\end{equation}

As a general result from Dupeux et al.\cite{Dupeux:2011}, the force per individual ratchet generally scales as
$f\sim\mu_v\frac{U}{h}R \lambda.$
Since the flow in the ratchet comes from the lubrication theory, $U\sim(h^2/\mu_v)(dP/dx)$, and taking into account the number of ratchets under the droplet, we obtain a general form of the  total propelling force
\begin{equation}
F\sim\mu_v\frac{k\Delta T}{L\rho_v}\left(\frac{R}{h_o}\right)^4\theta,
\label{generalForce}
\end{equation}
where $\theta=arctan(H/\lambda)$ is the ratchet angle with the horizontal (figure \ref{sketch}). To obtain such an expression it must be assumed that the aspect ratio is low enough to be taken as a small parameter; the limitations of such an approximation will be discussed below. For drops dominated by gravity $h_o=h_o^{(g)}$, eq. (\ref{hheavy}), and then the force propelling the droplet becomes
\begin{equation}
F^{(g)}\sim Mg \theta,
\label{Fheavy}
\end{equation}
where $M$ is the mass of the droplet, estimated as $M\simeq\rho_l R^2H_{drop}$.

We now come to the smaller, lightweight droplets, so-called capillary drops. For these drops, following the same arguments of Biance et al. \cite{Biance:2003}, the differences are:
(a) The contact radius can be estimated as $r_c\sim R^2/l_c$, where $l_c=\sqrt{\gamma/\rho_l g}$ is the capillary length for water, and
(b) the capillary pressure $\Delta P \sim \gamma/R$ becomes stronger that the pressure exerted by the weight of the droplet. Combining these two properties of capillary drops we can rewrite equations (\ref{evaporationrate}) and (\ref{lubrication}) to obtain the film thickness characteristic of capillary drops $h_o^{(c)}$,
\begin{equation}
h_o^{(c)}\sim\left(\frac{k\Delta T}{L\rho_v}\frac{\mu_v}{\gamma l_c}\right)^{1/4}R^{5/4}.
\label{hcapilar}
\end{equation}
Using this expression, we obtain the typical force propelling a capillary drop, which scales as
\begin{equation}
F^{(c)}\sim \gamma \frac{l_c^2}{R} \theta.
\label{Fcapilar}
\end{equation}
This result is not surprising if we realize that in both cases the pressure driving the flow under the droplet determines the scaling of the force driving the droplet movement.

\begin{figure}[h]
\includegraphics[width=0.5\textwidth]{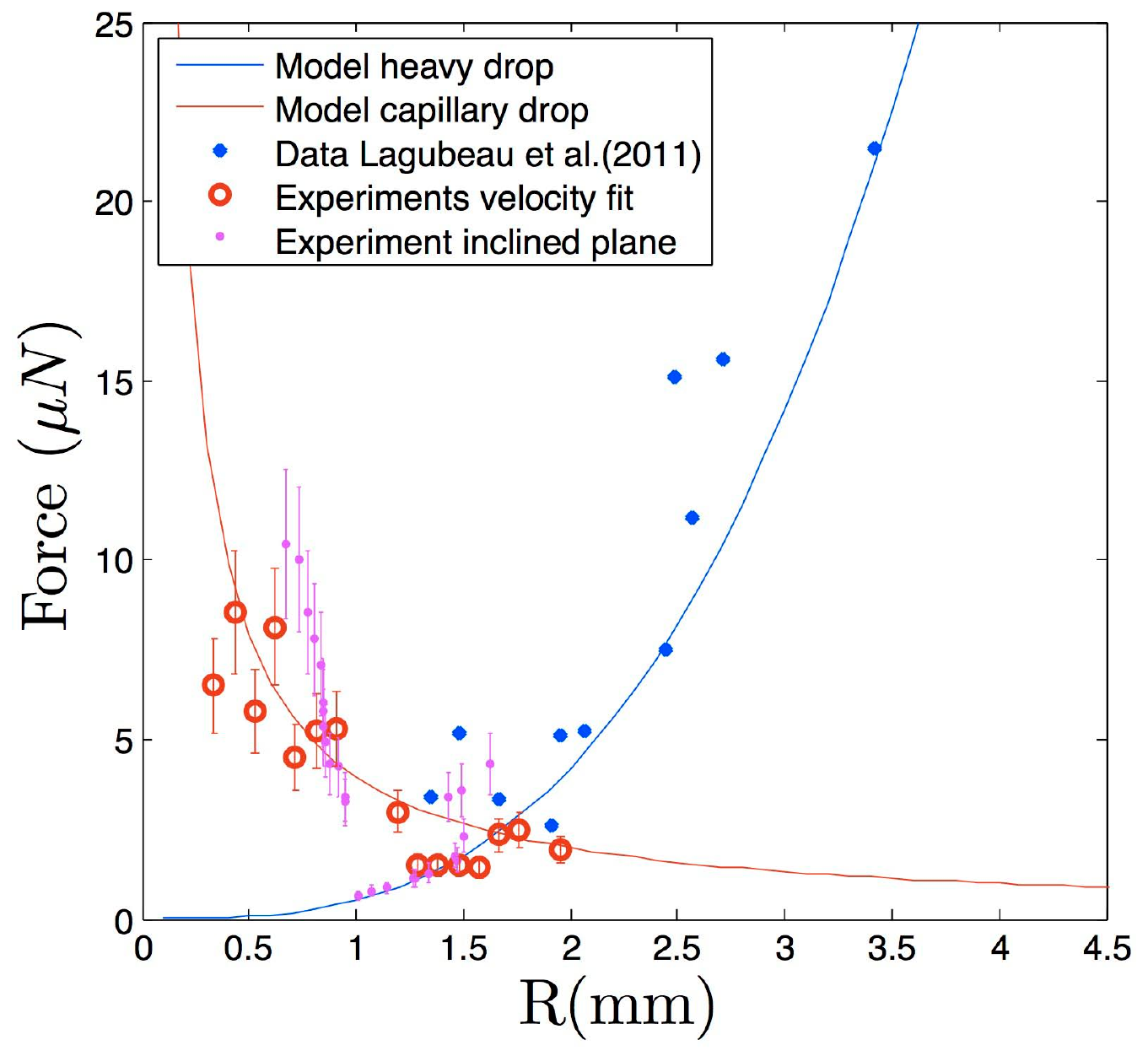}
\caption{Propelling force measurements and predictions from equations (\ref{Fheavy}) and (\ref{Fcapilar}) for capillary drops and for heavy drops. Results from Lagubeau et al. (2011) are also plotted for comparison.}
\label{FvsR1}
\end{figure}

In figure \ref{FvsR1} we compare both scalings with the measured force data described in previous sections. The missing prefactors in expressions (\ref{Fheavy}) and (\ref{Fcapilar}) have been chosen to fit the experimental data, resulting in $0.1$ for the heavy droplets and $0.33$ for the capillary drops. Heavy drops propulsion approximately scales with the droplet volume. The force decreases with the drop size until the droplet size becomes comparable with the capillary length for water. In this moment, both gravitational and capillary pressures are of the same order. As the droplet size decreases further, the force starts to increase. Smaller droplets have been measured, but they have the tendency to bounce off the substrate, losing contact with the vapor film. We conclude that the simple model by Dupeux et al. \cite{Dupeux:2011}, based on Biance et al. \cite{Biance:2003}, fits well with the measured data for both capillary and heavy drops.
The main source of data scattering are the unavoidable droplet oscillations in heavy drops and the mild bouncing in the capillary drops shown in the plot. These oscillations and bounces critically affect the contact area and the vapor film thickness and therefore a large natural scatter is found in the measured propelling forces.

One of the limitations of the simplified model is the assumption of a low ratchet aspect ratio (and therefore small $\theta$), which is not completely true in our case. This assumption is necessary for the lubrication approximation applied to the vapor film under the droplet, but is not fundamental for the mechanism. Certainly, a full analysis as that performed by Baier et al. \cite{baier} yields a more complex and stronger dependence of the propelling force with the ratchet angle $\theta$, but the dependence with the droplet size, which is our main concern in this work, remains intact. For the lubrication approximation to be valid, the Reynolds number associated to the vapor flow needs to be low enough. To calculate the maximum Reynolds number achievable in the micro-ratchet we assume maximum vapor velocities of 100 mm/s, vapor viscosity of $1.63\times10^{-5} Pa~s$, vapor density of $0.6 kg/m^3$, and a maximum ratchet depth of $50 \mu m$. A value of $Re\simeq 0.18$, which is low enough to justify the lubrication approximation.

\section{Bouncing capillary drops}

\begin{figure}[h*]
\includegraphics[width=0.75\textwidth]{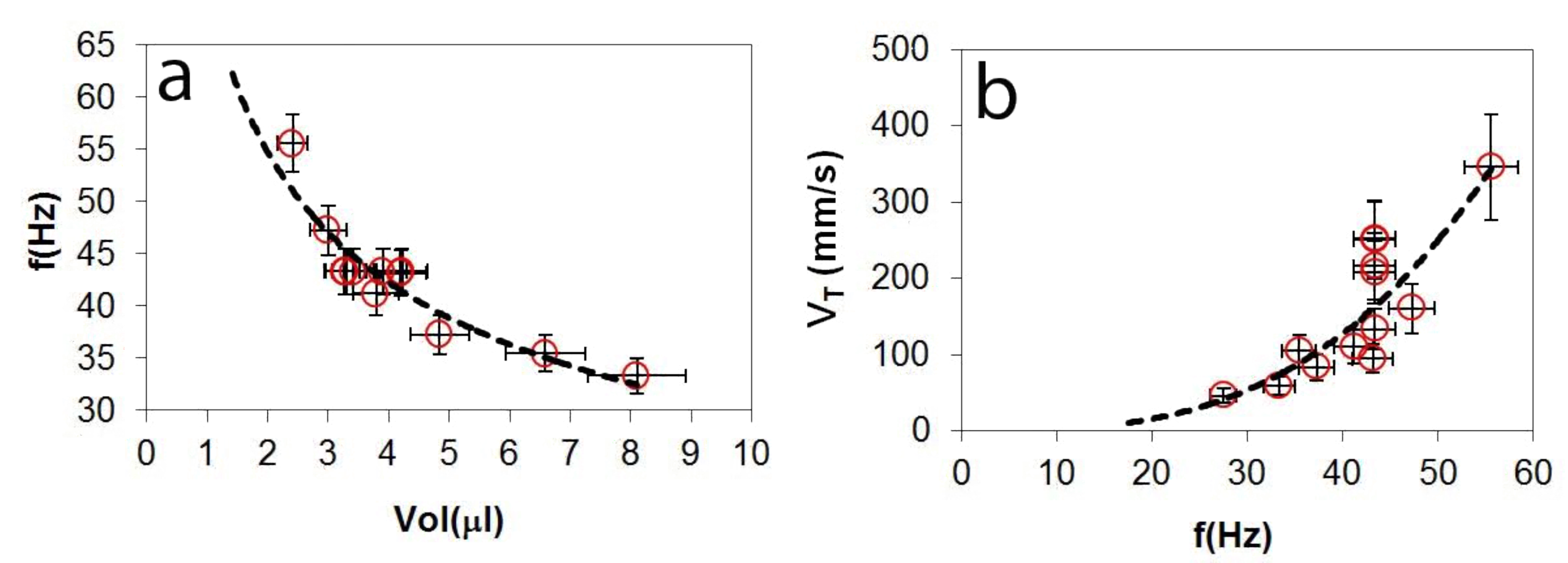}
\caption{(a) Bouncing frequency for different volumes of capillary drops. (b) Terminal velocity of bouncing drops for different frequencies. Only data from droplets with bouncing amplitudes comparable to their size are plotted here. The corresponding data for propelling force have not been employed in the analysis.}
\label{bouncers}
\end{figure}

Capillary drops tend to bounce off the surface upon contact with the overheated surface while being simultaneously propelled along by the micro-ratchets. Although understanding this peculiar behavior is not the main scope of the work, we performed measurements of the trajectories in order to characterize this phenomenon. For a single droplet, the typical maximum bouncing height observed is of the order of its own diameter. \footnote{For the propelling force measurements shown in previous sections, only those drops bouncing with a maximum vertical displacement below their own radius have been used.} The bouncing height decreases as the droplet is propelled forward, at different rates depending on the initial maximum height. The bouncing frequency is calculated by fitting a truncateated Fourier polynomial to the droplet trajectory, the Fourier transformed polynomial is then used to calculate the dominating frequency. The frequencies are calculated in lengths of 4 cm, where the typical amount of maxima observed (when the droplet reaches its maximum height) in a measurement is between 5 and 15.
In figure \ref{bouncers} we plot the typical bouncing frequencies for different capillary droplet volumes and velocities. We note that the bouncing frequency is inversely proportional to the droplet size, and the terminal velocity is directly proportional to the frequency. 
Regarding the latter, we have already observed and explained that the propelling force increases as the droplet size decreases for capillary droplets. Therefore, assuming that the origin of the vertical bouncing is not related with the force propelling the drop, we conclude that the higher velocities is unrelated with the bouncing effect.
Regarding the inverse dependence of the bouncing frequency with the droplet volume, a likely explanation (pointed out by one of the referees) can be found by taking the bouncing droplet as an harmonic oscillator or a mass on a spring, where the role of the spring would be taken by gravity. Therefore, the oscillating frequency scales as the inverse square root of the droplet mass, which would fit with the result found.

\section{Conclusions}

Our results show that the viscous mechanism proposed by Dupeux et al. \cite{Dupeux:2011} fits reasonably well with the experiments performed on micro-ratchets, not only for big droplets, but also for capillary droplets, in which the scaling of the force with the droplet radius is inverted. This regime can be achieved with laser-etched micro-ratchets, in which the size of the capillary droplets is at least ten times larger than the typical ratchet size. Alternative scalings, as that proposed initially by Lagubeau et al. \cite{lagubeau2011leidenfrost}, based on the rectification of the vapor ejected from the droplet, do not reproduce the scaling inversion of the force for smaller droplets observed in the experiments.
According to the analysis shown, the only geometrical parameter relevant to the droplet thrust is the ratchet aspect ratio and not its length or its depth independently. The results obtained by Ok et al.\cite{Ok} confirm that it is possible to achieve comparable results with ratchets of even a few hundreds of nanometers. It is indeed surprising that the proposed viscous mechanism is the dominant one from the millimetric scale down to the nanometric one. Recently, a completely different scaling has been proposed by W\"urger \cite{wurger2011}, based on the thermal creep velocity at the ratchet surface, i.e. temperature gradients that provoke vapor movement along the solid-gas interface. Such a mechanism would require extremely high temperature gradients along the metallic surface of the ratchet, and even with those, the velocity would be only concentrated in the first molecular layers close to the solid surface, while the liquid-vapor interface is at typical distances of $50~\mu m $. The model by W\"urger \cite{wurger2011} gives a similar scaling as that suggested by Lagubeau, but the force itself is 3 or 4 orders of magnitude smaller than the experimental data shown here, or that shown by both Lagubeau et al. \cite{lagubeau2011leidenfrost} and Dupeux et al. \cite{Dupeux:2011}. 
In this work we presented solid evidence to support the theory that the viscous mechanism is the primary driver of the phenomenon, even with micro-ratchets in the reported size. The heated ratchet is a relatively simple device to work with, but it is difficult to make accurate force measurements or to visualize the vapor flow in the ratchet. Therefore, in order to have more accurate data to test the proposed models, it would be necessary to produce different devices which allows for visualization of the vapor flow in asymmetric cavities.
The current knowledge on the thrust, necessary to move the droplets, and on the friction, necessary for their control, could bring interesting applications not only for the transport of droplets, but also for the controlled evaporation of colloidal or polymer droplets to create solid micro/nanoparticles or complex particles (capsules, janus...etc) in selected spots on special surfaces\cite{LFPreduct}.  
\begin{acknowledgements}
The authors acknowledge discussions with G. Dupeux, G. Lagubeau, D. Qu\'er\'e, and very specially we acknowledge the contributions from C. Clanet.
\end{acknowledgements}


\end{document}